\documentstyle[axodraw,epsf,sprocl]{article}

\def\NIMA{{\em Nucl. Instrum. Methods} A}
\def\NPA{{\em Nucl. Phys.} A}

\def\PRC{{\em Phys. Rev.} C}

\begin{document}

\title{SIMULTANEOUS HEAVY ION DISSOCIATION AT  
ULTRARELATIVISTIC ENERGIES}

\author{
I.A. PSHENICHNOV$^{1,2}$\footnote{e-mail: pshenichnov@nbi.dk},
 J.P. BONDORF$^{3}$, S. MASETTI$^{2}$, \\   
I.N. MISHUSTIN$^{3,4}$, A. VENTURA$^{2}$ \\
{\em $^1$ Institute for Nuclear Research, Russian Academy of Science,}\\
{\em 117312 Moscow, Russia}\\
{\em $^2$ Italian National Agency for New Technologies,}\\
{\em Energy and Environment, 40129 Bologna, Italy}\\
{\em $^3$ Niels Bohr Institute, DK-2100 Copenhagen, Denmark}\\
{\em $^4$ Kurchatov Institute, Russian Research Center,}\\
{\em 123182 Moscow, Russia}\\
}
\date{ }
\maketitle
\abstracts{
We study the simultaneous dissociation of heavy ultrarelativistic nuclei 
followed by the forward-backward neutron emission in 
peripheral collisions at colliders. The main contribution to this 
particular heavy-ion dissociation process, which can be used as a 
beam luminosity monitor, is expected to be due to the electromagnetic 
interaction. The Weizs\"{a}cker-Williams method is extended to the case of 
simultaneous excitation of collision partners which is simulated by 
the RELDIS code. A contribution to the dissociation cross section due 
to grazing nuclear interactions is estimated within the abrasion 
model and found to be relatively small.
}

\section{Single and mutual electromagnetic dissociation }
Let us consider a collision of heavy ultrarelativistic nuclei 
with the masses and charges $(A_1,Z_1)$ and $(A_2,Z_2)$ 
at the impact parameter $b$ 
exceeding the sum of nuclear radii, $b > R_1+R_2$.
According to the Weizs\"{a}cker-Williams (WW) method,
the impact of the Lorentz-boosted Coulomb field of the nucleus $A_1$ 
on the collision partner $A_2$ is treated as the absorption of  
equivalent photons~\cite{Pshenichnov2}. The mean number of photons absorbed 
by the nucleus $A_2$ in such collision is: 
\begin{equation}
m_{A_2}(b)=
\int N_{Z_1}(E_1 ,b)\sigma_{A_2}(E_1)dE_1 ,
\label{mb}
\end{equation}
\noindent where the spectrum of virtual photons, 
$N_{Z_1}(E_1 ,b)$~\cite{Pshenichnov2}, and 
the total photoabsorption cross section, 
$\sigma_{A_2}(E_1)$, are used. 
   
Assuming that the probability of multiphoton absorption is 
given by the Poisson 
distribution with the mean multiplicity $m_{A_2}(b)$,
one has the cross section for the single electromagnetic dissociation to a
given channel $i$~\cite{Pshenichnov2}:
\begin{equation}
\sigma_1(i)=2\pi\int\limits_{b_{c}}^{\infty} bdb P_{A_1}(b),\  
b_c=1.34\bigl( A_1^{1/3}+A_2^{1/3}-0.75(A_1^{-1/3}+A_2^{-1/3})\bigr),
\label{SS}
\end{equation}
where $b_c$ is in fm and the probability of dissociation at 
impact parameter $b$ is given by:  
\begin{equation}
P_{A_1}(b)=e^{-m_{A_1}(b)}\int  dE_1 N_{Z_1}(E_1,b)
\sigma_{A_2}(E_1)f_{A_2}(E_1,i).
\label{P1}
\end{equation}
Here $f_{A_2}(E_1,i)$ is the branching ratio for the considered channel $i$ in
the absorption of a photon with the energy $E_1$ on the nucleus $A_2$. 
These values are calculated by photonuclear reaction 
models~\cite{Chadwick,Iljinov} or taken from experiments.     

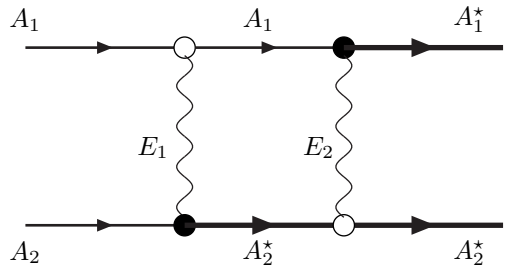
\begin{figure}
\begin{center} 
\vspace{1cm} 
\unitlength=1cm 
\begin{picture}(8.0,2.0) 
\SetWidth{1.0}
\ArrowLine(0,100)(60,100)  
\Text(0,3.9)[]{$A_1$} 
\SetWidth{0.5}  
\GCirc(60,100){4}{1.} 
\SetWidth{1.0}
\ArrowLine(64,100)(120,100)  
\Text(3.1,3.9)[]{$A_1$}   
\SetWidth{0.5}
\Photon(60,97)(60,36){3}{4} 
\Text(1.7,2.2)[]{$E_1$} 
\SetWidth{1.0}
\ArrowLine(0,33)(60,33)  
\Text(0,0.8)[]{$A_2$} 
\SetWidth{0.5}  
\GCirc(60,33){4}{0} 
\SetWidth{2.0}
\ArrowLine(60,33)(120,33)  
\Text(3.1,0.8)[]{$A_2^\star$}
\SetWidth{0.5}  
\GCirc(120,100){4}{0}
\SetWidth{2.0} 
\ArrowLine(120,33)(180,33)  
\Text(5.9,0.8)[]{$A_2^\star$}
\SetWidth{0.5}  
\GCirc(120,33){4}{1.} 
\Photon(120,97)(120,36){3}{4} 
\Text(3.9,2.2)[]{$E_2$}
\SetWidth{2.0}
\ArrowLine(120,100)(180,100)  
\Text(5.9,3.9)[]{$A_1^\star$}   
\end{picture} 
\end{center} 
\caption{ 
Mutual electromagnetic dissociation of relativistic nuclei. 
The open and closed circles denote the elastic and inelastic vertices,
respectively.  
\label{EMsecond}  
} 
\end{figure}

In the WW method, the graph for the mutual electromagnetic dissociation, 
Fig.~\ref{EMsecond},  
may be constructed from two graphs of the single 
dissociation by interchanging the roles of ``emitter'' and 
``absorber'' at the secondary photon exchange.
Such procedure is possible since the first emitted photon with 
$E_1\leq E_{max}\sim \gamma /R$ does not change essentially  
the total energy, $E_A=\gamma M_A$, of the emitting nucleus,
$E_{max}/E_A\approx 1/R M_A\sim 10^{-4}$, and there are no correlations
between the energies $E_1$ and $E_2$. In other words,
both photon exchanges may be considered as independent processes.
Moreover, at ultrarelativistic energies the collision time is much 
shorter then a typical deexcitation period when a nucleus loses its charge via
the proton emission or fission. 
It means that the equivalent photon spectrum from
the excited nucleus, $A_2^\star$, is equal to the spectrum from the nucleus 
in its ground state, $A_2$, see Fig.~\ref{EMsecond}. 

Therefore the cross section for the {\em mutual} electromagnetic
dissociation of the nuclei $A_1$ and $A_2$  to given channels $i$ 
and $j$ is given by:
\begin{equation}
\sigma_m(i,j)=2\pi\int\limits_{b_{c}}^{\infty} bdb 
P_{A_1}(b) P_{A_2}(b).
\label{SM}
\end{equation}
Substituting $P_{A}(b)$ for each of the nuclei one has:
\begin{equation}
\sigma_m(i,j)= \int\int dE_1dE_2 
{\cal N}_m(E_1,E_2) \sigma_{A_2}(E_1)\sigma_{A_1}(E_2) 
f_{A_2}(E_1,i) f_{A_1}(E_2,j),
\label{SMFULL}
\end{equation}
\noindent with the spectral function ${\cal N}_m(E_1,E_2)$ defined 
for the mutual dissociation:
\begin{equation}
{\cal N}_m(E_1,E_2)=2\pi\int\limits_{b_{c}}^{\infty} bdb 
e^{-2m(b)} N_{Z_1}(E_1,b) N_{Z_2}(E_2,b).  
\label{NM}
\end{equation}

\section{Nucleon removal in grazing nuclear collisions}
The cross section for the abrasion of $n$ neutrons and $z$ protons from the 
projectile $(A_1,Z_1)$ in a peripheral collision with 
the target $(A_2,Z_2)$ may be derived
from the Glauber multiple scattering theory~\cite{Hufner,Benesh}.
A simple parameterization exists for the single neutron removal cross
section~\cite{Benesh}: 
\begin{eqnarray}
\sigma_{nuc}(1n)=\frac{A_1-Z_1}{A_1}\sigma_G P_{esc},\ \ \ \ \
\sigma_G=2\pi\Bigl(b_c-\frac{\Delta b}{2}\Bigr)\Delta b,\ 
\nonumber
\label{NUC1N}
\end{eqnarray}
where $\Delta b\approx 0.5$ fm and $P_{esc}\approx 0.75$ is
the probability for a neutron to escape without suffering the 
interaction with a spectator fragment.
This can be extended to the case of mutual (1n,1n) emission:
\begin{eqnarray}
\sigma_{nuc}(1n,1n)=\Bigl(\frac{A_1-Z_1}{A_1}\Bigr)
                    \Bigl(\frac{A_2-Z_2}{A_2}\Bigr)
\sigma_G P_{esc}^2.
\label{NUC1N1N}
\end{eqnarray}

\section{Results and discussion}
\begin{figure}[ht]
\begin{centering}
\epsfxsize=0.7\textwidth
\epsffile{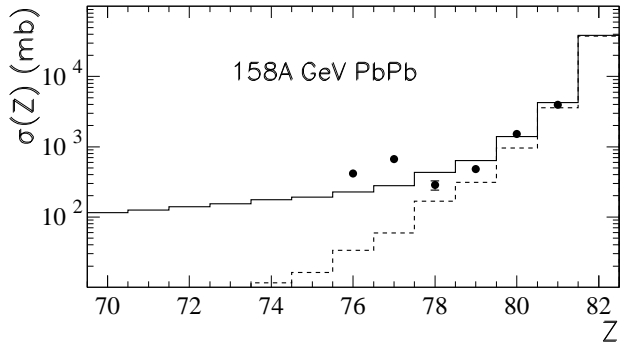}
\caption{Charge changing cross sections of 
$^{208}{\rm Pb}$ ions at CERN SPS energies.  
The solid and dashed-line histograms are the RELDIS code results 
with and without contribution of the abrasion
process, respectively.
The points are experimental data~\protect\cite{Dekhissi}.}
\label{fig:zdis}
\end{centering}
\end{figure}

The results of the abrasion model for the charge changing cross 
sections of the
{\em single} dissociation of 158A GeV $^{208}{\rm Pb}$ ions are 
shown in Fig.~\ref{fig:zdis} in addition to the electromagnetic 
contribution calculated by the RELDIS code. 
As one can see, the electromagnetic contribution dominates for
few nucleon removal process. 
The interaction of knocked out nucleons with spectators 
and spectator de-excitation process itself 
were neglected in this version of the abrasion model.
However, good agreement with the experimental data~\cite{Dekhissi} is found.

After this verification the model can be extrapolated to the energies of
RHIC and LHC heavy-ion colliders. There is a proposal to use the   
simultaneous neutron emission for beam luminosity 
monitoring via the correlated registration of forward neutrons in zero degree
calorimeters at RHIC~\cite{Baltz2}.
The model predicts $\sigma_m(1n,1n)\approx 750$ and 970 mb for AuAu and 
PbPb collisions at RHIC and LHC, respectively, 
for the correlated single neutron emission.  
We found that an important part of the dissociation events 
leading to the single neutron emission is accompanied by the emission 
of charged particles and nuclear fragments.
Such events are due to the equivalent photon absorption 
above the Giant Resonance region. For grazing nuclear collisions 
Eq.~(\ref{NUC1N1N}) gives  
$\sigma_{nuc}(1n,1n)\approx 100$ mb.
Our detailed consideration of the mutual nuclear dissociation is in
progress.

We are grateful to A.S.~Botvina, G.~Dellacasa, J.J.~Gaardh\o{j}e,
G.~Giacomelli, A.B.~Kurepin and S.~White for useful discussions. 
I.A.P. and I.N.M. are indebted to the Organizing Committee 
of Bologna 2000 Conference  for the kind hospitality and 
financial support.

\end{document}